\documentstyle[11pt,newpasp,twoside,epsf]{article}
\begin{document}

\title{Young Circumstellar Disks and Their Evolution: A Review }
\author{Lynne A.\ Hillenbrand}
\affil{California Institute of Technology; MC 105-24; Pasadena, CA 91125}

\begin{abstract}
A detailed understanding of the physics
of star and planet formation requires
study of individual objects as well as statistical 
assessment of global properties and evolutionary trends.
Observational investigations of circumstellar material surrounding 
young stars have matured to the point that both 
spectral energy distributions sampled over more than
four decades in wavelength and spatially resolved images 
or interferometric visibilities at limited optical/infrared 
and sub-/millimeter wavelengths are becoming available,  
though for few individual objects at present.  Data on
star/disk systems combined with continuing surveys 
for exo-solar planets themselves will lead to constraints on
the likelihood and frequency of solar system formation.
An overarching goal of these pursuits is to connect
what is observed elsewhere with the history of our own solar solar system,
and hence enhance our appreciation of the uniqueness -- or lack thereof -- 
of the human circumstance.
\end{abstract}

\section{Introduction}

Several decades of theoretical and observational work has led to the following
paradigm for star formation.  Nearly free-fall 
collapse of an initially slowly rotating ($\Omega < 10^{-13}$ s$^{-1}$), 
cold ($T = 10-20 K$), dense ($n\sim10^5$ cm$^{-3}$) molecular cloud 
core leads to formation of a hydrostatic proto-star surrounded by a 
centrifugally supported circumstellar disk and remnant infalling envelope.  
The main infall/accretion phase during which most of the stellar mass is 
acquired is heavily obscured (A$_V > 10^3$ mag) and relatively
short ($<10^5$ yr; Shu et al. 1987; Boss \& Hartmann 2001).
A self-luminous source emerges from within the shroud of its parent molecular
core and becomes visible (A$_V <$ 5-10 mag)
at near-infrared and optical wavelengths as a
pre-main sequence star contracting along a Hayashi track 
(in the most likely case of a low-mass object) in the HR diagram
(see, e.g. Palla, 2001 for a review).
The system evolves through the final stages of infall and then
disk accretion, which can last up to $\sim$10$^7$ yr in some cases 
(Muzerolle et al. 2000; Alencar \& Batalha 2002).
Competition between inward accretion and the simultaneously operating 
process of outflow along the magnetic/rotation axis affects both the mass 
and the angular momentum of the forming star, as well as the amount of 
raw material left in the disk for possible formation of planetary systems.
The disks dissipate on times scales as short as $<$10$^6$ years, in some cases,
and as long as $>$10$^7$ years, in a minority of cases,
due to a combination of accretion onto the central star, ablation
by the star's radiation field or wind, and/or assembly of smaller particles
into larger bodies.  
For solar-type stars, the ultimate result in at least 
10\% and perhaps as many as 50\% of cases is a mature solar system.

Over the past 5-7 years we have witnessed 1) dramatic confirmation
through direct imaging of many aspects of this picture,
which was established previously using indirect arguments, and
2) the enlightening discovery of fully formed
exo-solar planets and planetary systems.  
The remainder of this introduction outlines the existing observational evidence
and theoretical support that underpins our still largely schematic 
understanding of planet formation as a concomitant to star formation, 
illustrated in Figure 1.  We then discuss in some detail current constraints 
on the time scales for circumstellar disk dissipation.
Mannings, Boss, \& Russell (2000) provide an extensive set of reviews
on many topics relevant to the ideas expressed herein; see also
Podosek \& Cassen (1994) for a briefer look at theory, observations,
and solar system constraints on planet formation timescales.

\subsection{Pre-Collapse: The Structure of Molecular Cloud Cores}
Before initiation of star formation, molecular gas organizes itself
on large scales (10s-100s of pc; $>$10$^4$-10$^5$ M$_\odot$) into clouds, sheets, 
and filaments, and on smaller scales ($<$0.1-1 pc; $<$10-10$^3$ M$_\odot$),
into individual protostellar and proto-cluster cores.
The structure of pre-collapse molecular cores has been inferred from 
observations of marginally resolved thermal emission at 
far-infrared through millimeter wavelengths.  These data 
can be fit with power-law density distributions $\rho(r) \propto r^{-\alpha}$ 
with $\alpha$ closer to 0.5-1 (flatter profiles) in the inner regions and 2-4 
(steeper) in the outer regions (Ward-Thompson et al. 1999, 1994; 
Andre et al. 1996).  Tafalla et al. (2002) adopt the hybrid functional form 
$\rho(r) = \rho_o/(1+(r/r_o)^{\alpha})$. 
Disagreement remains, however, concerning the need for deviation 
from the isothermal prediction, $\alpha$ = 2 
(Shirley et al. 2002, 2000; Evans et al. 2001), and whether there is
any consistency with the free-fall prediction, $\alpha$ = 1.5. 
Recently, Alves et al. (2001), Harvey et al. (2001), 
and Racca et al. (2002) have probed the structure
of high column density material at even higher spatial resolution
using optical/infrared photometry of background stars
to measure surface density profiles of several Bok globules.
The observations are well-described by more sophisticated (compared to 
simple power-law functions) Bonner-Ebert spheres. 
A number of near-critical, that is, on the verge of or just having begun 
collapse, cloud cores have been identified.  

Evidence for transition from molecular cores to proto-stars comes 
from observations of characteristically asymmetric line profiles in CO, CS, 
and other high density traces (see reviews by Myers et al. 2000
and Evans 1999) which demonstrate the earliest stages of gravitational 
collapse and infall.  Indeed, one source studied in this manner, B335, 
was also found to be an unstable Bonner-Ebert sphere using the 
extincted starlight techniques just discussed.  The typical size of the region 
inferrred to participate in the infall is $<$0.1 pc.

\subsection{Early Stages: Spherical Infall and Disk Accretion.}

Following the main collapse phase, a proto-star emerges.  
Proto-stars are usually defined according to criteria regarding the
the fractional infrared through sub-/millimeter luminosity, the wavelength of
peak emission in the spectral energy distribution, the spectral index,
or ideally the mass in the central object compared to mass 
in the infalling envelope (see, e.g. Andre et al. 2000). 
One of their observational signposts are strong molecular outflows thought
to be driven by highly collimated jets.
Hundreds of protostellar objects have been cataloged within a few kpc
of the Sun.  Modelling of the nearest of these, in molecular clouds
$\sim$150 pc distant, has shown that their spectral energy distributions 
are generally well-matched at long wavelengths by simple radiative transfer of 
released gravitational energy through
spherically symmetric infalling circumstellar envelopes. At short wavelengths
(2-3 $\mu$m), however, there is suggestion based on flux excesses
above the model predictions (Kenyon et al. 1993a) as well as
spatial and polarization information (Kenyon et al. 1993b
and references therein), that the optical depths to the central sources
are far less than the pure-spherical infall picture predicts.
An extended but flattened dust geometry is suggested, and confirmed by 
imaging (e.g. millimeter, Ohashi et al. 1996;
HST/WFPC, Stapelfeldt et al. 1999;
HST/NICMOS, Cotera et al. 2001, Padgett et al. 1999)
which shows disk-like
dust lanes separating bipolar cometary-shaped nebulae.
The simultaneous consideration of the spectral energy distribution and 
spatially resolved images at several wavelengths (Padgett et al. 2003) 
enables detailed modelling which leads to physical parameter estimation.
The composite observations of these protostellar systems are best
explained if there are inner cavities in the infalling envelopes,
which allow short-wavelength photons from some combination of the central
object, the inner accretion disk, and the base of the jet/outflow region
emitted towards the poles of the dust envelope to scatter and escape into
the line-of-sight of the observer.  Although some of these scattered light
images have been obtained with HST and Keck, the majority of 
protostars in the nearest molecular clouds are fainter than I=25
and will require NGST/CELT for detailed study in the optical/infrared as well
as SOFIA/ALMA for mid-infrared/sub-millimeter work at high spatial resolution.

After the envelope has fully settled into the existing disk, the central
star becomes an optically visible T-Tauri (lower masses) or Herbig Ae/Be 
(higher masses) object.  Due to the short time scales involved, observation 
of the accretion disk phase around even higher mass early-B and O-type stars 
is rare.
Global disk heating is provided from two main sources: dissipation of viscous 
energy due to accretion, and radiant heating of the disk surface
by the central stellar object.
Additional heating processes such as chromospheric flaring events 
and x-ray ionization and flaring, may also be important.  
Observable signatures at this stage include mid-infrared to sub-/millimeter
emission from dust at a range of temperatures, near-infrared 
emission from hot dust and gas in the inner disk, and a variety 
of optical and ultraviolet emission lines (as well as continuum excess) 
due to accretion of material from the inner disk directly onto the star.  
The strongest evidence 
for accretion comes from observation at high spectral resolution of the hot, 
blue excess above the expected photospheric spectrum, and detailed modelling 
of atomic emission line profiles.  ``Typical'' low-mass
($\sim$0.5 M$_\odot$) young (1-3 Myr) stars in the well-studied Taurus-Auriga
molecular cloud have accretion rates spanning a wide range, from 
$\sim$10$^{-9}$-10$^{-7}$ M$_\odot$yr$^{-1}$ (e.g. Valenti et al. 1993,
Hartigan et al. 1995, Gullbring et al. 1998, White \& Ghez 2001) with
modest evidence for a direct correlation with stellar mass 
(e.g. Hillenbrand et al. 1992; Rebull et al. 2002, 2000; White \& Basri 2003).

\subsection{Middle Stages: Termination of Accretion and Planet Building}

As was true for the protostellar stage, much excitement over the past 
several years has come from direct imaging of later disk evolutionary phases. 
In the Orion nebula, for example, O'Dell  et al. (1993) and 
McCaughrean \& O'Dell (1996) presented 
stunning images of so-called silhouette disks which provided visual evidence
of flattened disk-like geometry associated with young stars
for the first time shortward of millimeter wavelengths.
Optical/infrared imaging of scattered light and thermally emitting 
disks surrounding optically visible stars has been provided by HST  
(e.g. Schneider et al. 2000; Krist et al. 2000;
Weinberger et al. 2002; Grady et al. 2001)
and ground-based AO (e.g. Roddier et al. 1996, Close et al. 1998).  Resolved
structure has also been found in interferometric studies
(e.g. Akeson et al. 2000; Tuthill et al. 2001; Millan-Gabet et al. 2001; 
Hinz et al. 2001; Eisner et al. 2003).  These images/visibilities
provide information on orientation and on disk size vs. wavelength 
which, in combination with spectral energy distributions, 
has led to a new class of models for Herbig Ae/Be stars in particular 
(Dullemond et al. 2003).

For infalling protostellar envelopes as well as for active accretion and 
more passive ``reprocessing" disks,
modelling of optical through millimeter spectral energy distributions 
using radiative transfer techniques is hindered by the many
degeneracies between: dust composition, grain size distribution, 
radial/vertical structure and scale heights, and overall geometry including
system inclination.  Longer wavelengths measure cooler
temperatures which generally correspond to larger radii,
though spectral energy distribution details will depend on
the specifics of the radiative transfer, including scattering.  
An active debate at present concerns the three-dimensional geometry 
of such disks, in particular the existence and size of inner disk 
magnetospheric holes, the degree of outer disk flaring 
($\alpha$ in the expression $(h/R)^\alpha$ where 
h is the vertical height and R is the corresponding disk radius),
the evidence for settling of material within the disk
vertically to the mid-plane which may be a critical early step in
planet formation, and the opening of gaps in disks as
a consequence of planet formation.  Increase in surface
brightness sensitivity through high contrast direct imaging and
interferometric techniques should resolve large-scale
three-dimensional structure or place clear limits on 
the degree of disk flaring. 
Observations of holes, gaps, truncations, warps, and non-axisymmetric structure,
all plausibly induced by planet-disk interactions
(e.g. Calvet et al. 2002, Wolf et al. 2002, Ozernoy et al. 2001), 
may offer the first direct evidence of ongoing or recent planet formation.

\subsection{Late Stages: Debris Disks}

Beyond the ``primordial" disk stage discussed so far, in which the disks are 
still undergoing initial dissipation of the material out of which they and their
central star formed, is the ``secondary" or so-called debris disk stage in which
second-generation dust is produced in collisions between 
planetesimals that are themselves stirred by gravitational interactions 
with larger planets, and by comets.  
A working definition of a debris disk (in the absence of gas) is one in which 
the survival time of the dust -- given drag into the star via the Poynting-Robertson
effect and/or radiation blowout -- is much shorter ($\sim$10$^5$ yr is a typical
number) than the stellar age.  Our own zodiacal dust is an example in which 
the asteroid belt and Edgeworth-Kuiper Belt are the source regions for the 
parent bodies which produce dust (e.g. Stern 1996) 
due to perturbation of the smaller bodies in these belts
by our gas giant planets, primarily Jupiter and Neptune.

Debris disks can be distinguished from primordial disks in the following ways: 
They are low mass ($<$0.01 M$_{\earth}$) and gas poor ($<$10\%), 
the dust is located mainly at large radii and hence cold, 
the system dynamics are dominated by radiation pressure /collisions,
and most significantly, the grains are in the process of being destroyed.
Primordial disks on the other hand, are more massive (0.01 - 0.1 M$_\odot$)
consisting of both dust and gas, have material distributed over a large radial 
range with a corresponding range of temperatures from warm to cold, have 
system dynamics dominated by a more massive gas component perhaps in
keplerian rotation, and finally, are undergoing a dominant process of
grain sticking/growth.  It can be quite difficult
to distinguish between these two scenarios (primordial and secondary)
for any given observed system, especially at younger ages, say in the 3-30 Myr 
range.

Debris disks were discovered by IRAS (Aumann et al. 1984) via their thermal 
emission and later spatially resolved 
in short-wavelength scattered light and long-wavelength thermal emission 
from the ground (e.g. Smith \& Terrile 1984; Holland et al. 1998) and
from space (e.g. Weinberger et al. 1999, Heap et al. 2000; Heinrichsen 
et al. 1998).  The few debris disks 
which have been resolved appear asymmetric, radially and/or azimuthally, and
exhibit morphologies consistent with those predicted from the dynamical
response of the disk to an embedded planet or planets
(e.g. Wyatt et al. 1999; Liou \& Zook 1999; Moro-Martin \& Malhotra 2002).  
The exo-solar debris disks 
(see Lagrange et al. 2000 and Backman \& Paresce 1993 for reviews)
along with studies of the zodiacal dust and asteroid+Kuiper belts in our own
solar system (see Luu \& Jewitt 2002 for a review) have suggested a close
relationship between late-stage debris disks and the presence of
planetary systems.

Giant planets are now known to orbit within $\sim$3 AU of $\sim$6\% of 
solar-type stars in the solar neighborhood (Marcy et al. 2000), 
with increasing evidence for commonality of multiple planet systems 
(Fischer et al., this volume; Marcy et al. 2003 website).  The bulk properties 
of the planets discovered to date are {\it dis}-similar to those
in our own solar system, in that semi-major axes typically are much smaller and
eccentricities typically much higher; nevertheless, several near-analogs 
of Jupiter and/or Saturn have been found already with more surely 
on the horizon.
The mere existence of exo-solar planets has substantially increased interest
in the connection between the disks commonly seen around
young stars, and solar systems such as our own.

\section{Potential for Planet Formation in Young Circumstellar Disks}

Let us now backtrack to the middle stages (adolescence, if you will) 
to discuss in more detail the expectations and observational evidence
concerning primordial disk dissipation.
The basic schematic from proto-stars to planets as outlined
in the introduction is well accepted, though the details of evolution, 
including relevant time scales, are still quite poorly constrained. 
There is now overwhelming observational evidence that young circumstellar 
dust and gas disks exist.  Questions then arising -- many of which likely
can not be answered this decade -- include: 
How, and on what time scales, does the gas and dust found
near ubiquitously around young solar-type stars evolve? Do disks last long 
enough to form planets?  When, where, and how 
frequently do planets form in circumstellar disks? How do forming
planetary systems evolve dynamically? 
Are there variations in solar system birth and evolution 
with mass or other properties of the parent star,
or with local circumstellar environment?
What is the range in diversity of stable planetary system 
architectures?   How unique is our own solar system? How frequent are 
habitable planets?  

\subsection{Bulk Properties Consistent with Pre-Planetary Solar Nebula}

We consider first the likelihood that young dust and gas disks 
are in fact proto-planetary. The raw material of planetary embryos, 
earth-like rocks, and jupiter-like gas giants is abundant. 
Whether any individual disk {\it will} form planets 
is of course unknowable.  But that, in the mean, they are at least
{\it capable} of forming planetary systems similar to our is almost a certainty,
as evidenced from observed disk sizes, masses, and composition/chemistry.

Disks around young stars were spatially resolved for the first time at millimeter
wavelengths (e.g. Sargent \& Beckwith 1987) which measure cold gas and dust 
in the outer disk regions. Unequal axial ratios, combined with implied dust
masses large enough that the central stars should not be optically visible 
if the dust geometry is spherical, stood as the strongest evidence for close 
to a decade that these are in fact disks and hence potential proto-planetary
systems.  Continued interferometric work (e.g. Lay et al. 1994,
Dutrey et al. 1996, Duvert et al. 2000; Kitamura et al. 2003) 
suggests that disk diameters -- {\it in instances where spatially resolved 
images are in fact obtained} -- range from 
70-700 AU and are even as large as 2000 AU in some cases.  
These size estimates are consistent with those inferred
from the optical/near-infrared scattered light images discussed above, and
in the typical case are comparable to or larger than our solar system.
Surface density
profiles, e.g. simple power-laws with $\Sigma(r) \propto r^{-p}$ or viscous disk
``similarity solutions'' with with $\Sigma(r) \propto r^{-p} e^{-r^{(2-p)}}$ 
have suggested a wide range in the value of p (0-1.5 for the power-law case).  
Furthermore, 
kinematic models of spatially resolved CO emission suggest Keplerian rotation 
(e.g. Koerner et al. 1993; Mannings et al. 1997; Simon et al. 2000; 
Koerner \& Sargent 2003).

Disk masses are derived from optically thin millimeter flux and an
adopted opacity-wavelength relationship which leads to uncertainties of 
factors of 5-10.  Under common assumptions, however, calculated dust masses 
range from 10$^{-4.5}$ to 10$^{-3}$ M$_\odot$ (e.g. Beckwith et al. 1990).
Making the further assumption that the dust:gas ratio by mass is 
unaltered from the canonical interstellar value of 1:100, 
total disk masses average around 0.02 M$_\odot$, or about the
Minimum Mass Solar Nebula (Kusaka et al. 1970;
Weidenschilling 1977) which is the reconstitution
of present-day solar system mass and composition to solar consistency.
It should be stressed that, as was true for the spatially resolved emission,
{\it detection at all of millimeter flux}
is made amidst an increasingly large 
number of upper limits, and so the true ``mean mass" is even lower than
quoted above.

The composition of both young primordial and older debris disks has been
shown to resemble that of solar system comets. Ground-based 
10 and 20 $\mu$m work on brighter sources 
(e.g. Hanner et al. 1995, 1998; Sitko et al. 1999) and
ISO 2-30 $\mu$m spectroscopy (e.g. Meeus et al. 2001; Bouwman et al. 2001)
have revealed an impressive suite of solid state (and PAH) dust features.
Mineralogical details of the dust are modelled on a case-by-case 
basis due to cosmic variance, but the mean composition 
appears to be $\sim$70-80\% amorphous magnesium-rich olivines,
$\sim$1-10\% crystalline forsterite, $\sim$10-15\% carbons, 
$\sim$3-5\% irons, and other trace components such as silicas.
In particular, crystallinity is advocated in $\sim$10\% of sources.
Studies of gas chemistry (see review by van Dishoeck \& Blake 1998;
also Aikawa et al. 2002) probe evidence for the building blocks of life. 
In the disk, temperature increases from the mid-plane outward
while density decreases outward, leading to
a strong dependence of fractionization on radius and height in the disk.

In summary, the observed sizes, masses, and dust/gas chemical composition
of young disks are all consistent with solar nebula estimates. The
mean disk properties are, however, still biased by 
detection limits and selection effects.

\subsection{Expected Evolutionary Time Scales}

We draw upon the ``ground truth" provided by our own solar system,
and also look to theory for guidance regarding the expected time scales 
for disk evolution.

At present there are three basic classes of models for planet formation.
In the accretion model
(e.g. Pollack 1996; Lissauer 1993; Weidenschilling \& Cuzzi 1993),
a multi-stage accumulation process involving dust settling and sticking
leads to slow and then run-away accretion of solids over 1-10 Myr,
followed by slow and then run-away accumulation of gas over the next
10-100 Myr.  The time scales are highly dependent on parameters such as the 
initial disk surface density; hence the range of disk lifetimes is not a 
firm predictable though gas disk lifetimes are expected to be longer than
dust disk lifetimes.  A hybrid between these pure accretion models and 
the pure fragmentation models discussed next is the theory 
(e.g. Youdin \& Shu 2002; Goldreich \& Ward 1973) 
that initial small particle sticking and settling to near the disk midplane
is followed by gravitational instability leading to planetary core contraction,
which may lead to further solid and gas accumulation.  
Planet formation time scales in this scenario 
are expected to be shorter than in the pure accretion models.
Pure fragmentation models (e.g. Boss 1998; Bodenheimer 1985) 
by contrast, advocate that giant planet formation is a relatively rapid 
($<<$1 Myr), direct effect in massive, gravitationally unstable 
gas disks.  This theory predicts short disk lifetimes and possibly even
early-stage radial velocity signatures of planetary-mass perturbers 
in the spectra of parent stars.
Finally, the capture theory of planet origins (e.g. Woolfson \& Oxley 2000) 
is still discussed.  In the mainstream theories of planet formation, 
possible co-existence of newly formed planets
and the accretion disk also leads to considerations of orbital migration
(e.g. Lin et al. 2000; Trilling et al. 1998), in which final orbital 
parameters, in particular the semi-major axis and eccentricity,
are different from those that obtain close to the epoch of planet formation.  

In our own solar system, meteoritic
evidence concerning survival time of the solar nebula suggests ``several
Myr'' as the relevant evolutionary time scale. 
Studies, especially those
concerning extinct radionuclides, support this time span for
initial accretion, differentiation, and core formation
(see e.g. review by Wadhwa \& Russell 2000). 
It should be emphasized that although dispersal of the solar {\it nebula}
may occur quickly, the total duration over which inner planet 
formation was completed in fact approached 50-100 Myr.

As we will describe, these limited constraints from theory and from
our own solar system are consistent with the equally vague precision 
with which disk lifetimes can be inferred from observations 
of potential planetary systems now in the making.

\subsection{Measures of Disk Evolution}

Resolved disk images as discussed in the introduction certainly 
have led to a wider appreciation of the convincing case for 
proto-planetary disks; however, the reality is that 
few such images exist at present. We still rely for the most part
on ``indirect" measurements such 
as broadband photometry and spectral energy distributions, as well as
high resolution optical and near-infrared
spectroscopy to study disks and disk accretion.
Young circumstellar disks can be depleted of material via several mechanisms.
These include: accretion of their dusty and gaseous material onto 
the central star, ``excretion" of material into an outflow 
generated close to the star in some models (Shu et al. 1994; Hirose et al.
1997) and further out in the disk in others (Konigl \& Pudritz 2000), 
ablation by the stellar ultraviolet radiation field or mechanical wind
(Shu et al. 1993) and finally, decrease of opacity due -- in the standard model --
to growth proceeding from insterstellar-like dust grains to rocky planetesimals.  
The potential influence of the environment in which normal disk evolution 
proceeds should not 
be ignored.  Location in a high stellar density region or in proximity to
a massive star may also accelerate disk evolution (see review by Hollenbach 
et al. 2000).

What can we hope to measure as primordial dust and gas disks evolve
through the planet building epoch into secondary/debris disks?
Possible diagnostics as a function of age include:
\begin{list}{$\bullet$}{\itemsep=-0.05truein}
\item evolution of disk geometry in direct imaging data 
\item evolution of the spectral energy distribution 
\item decay in measured disk accretion rates
\item increase in mean grain size and evidence for chemical evolution
\item decrease in dust mass
\item decrease in gas mass 
\item late stage dust regeneration signifying transition to a debris disk.  
\end{list}
Some, albeit in most cases limited, evidence for at least modest evolution 
from primordial disk conditions exists in all of these areas.  
In what follows we focus on evolution of spectral energy distributions.
In order to establish trends, robust statistics are needed from the youngest
ages characteristic of star-forming regions still associated
with molecular gas ($<$1-2 Myr),
through the entire period of gas giant and terrestrial
planet formation ($\sim$100 Myr for our own solar system), as depicted
in Figure 1.

\section{Disk Evolution in Young Stellar Clusters and Field Stars}

Young star clusters have a number of attributes that are attractive
for statistical studies of circumstellar properties, 
namely the relatively uniform 
distance, age, and chemical composition of their members. Clusters can therefore
provide the samples required to compare properties such as mean and dispersion 
in disk lifetimes as a function of stellar mass (within a cluster) 
and as a function of stellar age or chemical composition (between clusters). 
Known targets for this kind of work can be segregated
into the following coarse age groups:
$<$1 Myr (partially embedded star forming regions), 
1-3 Myr (optically revealed stellar populations still associated with
molecular gas), 10-15 Myr (association members in gas-poor 
``fossil" star-forming regions; N.B. this
is a particularly sparsely-populated age range at present), 
55 Myr, 90 Myr, and 120 Myr (the ages of nearest open clusters).
Stars in the 5-50 Myr age range are extremely hard to identify since they
stand out from much older field star populations only with detailed 
observations (not, e.g., in wide-field photometric surveys).
They may be revealed through such signatures of youth as common proper motion
with young groups, enhanced
Li I absorption, Ca II H\&K core emission, and x-ray activity.
In fact, finding stars in this age range should be relatively easy due to our 
circumstance in the Galaxy near a ring of moderately recent star formation 
(``Gould's Belt"); yet current samples of 5-50 Myr old stars
number only in the tens.

We now summarize the evidence for disk evolution over $<$1-100 Myr time scales
using available cluster, association, and field star samples.   As a caveat 
we note that the poor precision of stellar ages is still a major contributor to 
the uncertainty in derived time scales.

\subsection{Inner Accretion Disks}

It has been adequately demonstrated in the literature that there is an 
empirical connection between near-infrared (1-3 $\mu$m) flux excess and 
spectroscopic signatures of accretion directly onto the star,
as well as a connection between accretion and outflow 
(e.g. Hartigan et al. 1995).  We therefore utilize measured emission 
above expected photospheric values in the near-infrared 
to infer presence of an accretion disk. 
These wavelengths are sensitive to hot dust and gas in the inner disk, 
$<$0.05-0.1 AU, where the geometry of the emitting region may in fact be 
quite complex (see e.g. Mahdavi \& Kenyon 1998).

To calculate the color excess due to the disk, one must derive and subtract 
from the observed color the 
contributions from foreground/circumstellar extinction 
and from the stellar photosphere in order to arrive at an intrinsic excess, e.g.
$\Delta(H-K) = (H-K)_{observed} - (H-K)_{reddening} - (H-K)_{photosphere}$.
The information required for assessing $\Delta$(H-K) includes
a spectral type (for intrinsic stellar color and bolometric correction
determination), optical photometry (for dereddening and locating stars
on the HR diagram, assuming known distance), 
and infrared photometry (for measurement of disk ``strength").
The value of the infrared excess is affected by stellar properties
(mass, radius) and disk properties (accretion rate, inclination,
geometry); see Meyer et al. 1997 and Hillenbrand et al. 1998 for detailed
discussion.  Masses and ages are 
inferred via comparison to models of
pre-main sequence evolutionary tracks, which themselves carry uncertainties of
20-100\% between authors over certain mass and age ranges
(see comprehensive discussion in Baraffe et al. 2002).

Observationally, our best effort at measuring the evolution of inner
circumstellar accretion disks is represented in Figure 2, 
produced from a sample of $\sim$3000 stars 
located $\sim$50-800 pc from the Sun 
with sufficient information for calculation of $\Delta(H-K)$. 
There are several important points made by these plots. 
First, although individual stars (left panel of Figure)
in nearby star forming regions can appear to have ages $<$1 Myr, not a single
cluster or association (right panel of Figure) has median age significantly 
$<$1 Myr.  This is in part a selection effect since the requirements for
inclusion in our sample generally preclude the presence of proto-stars
and transitional objects.  Second,
even at the earliest evolutionary stages at which stars can be located in 
the HR diagram, the optically thick inner disk fraction does not approach unity. 
This may be influenced by the same selection effect, though indicates strongly
that {\it some} disk evolution does happen very early on for {\it some} stars,
before they become optically visible. Third,
beyond 1 Myr of age, where samples are more representative of stellar
populations as a whole (if not close to complete for most of the regions represented), 
there is a steady decline with time in the fraction of stars showing 
near-infrared excess emission (i.e. optically thick inner disks), 
as well as large scatter at any given age. 
The conversion of this Figure into a frequency distribution of accretion
disk lifetimes is discussed in Hillenbrand, Meyer, \& Carpenter (2003). 
Fourth, the median lifetime of inner optically thick accretion disks based 
on assessment of modern data may be as short as 2-3 Myr.  Comparison of the 
left and right panels of Figure 2 leads us to the conclusion that 
there is undoubtedly scatter of individual young stars into older age bins, 
which affects the appearance of the
left panel substantially more than that of the right panel.  The left panel
may thus be considered an upper limit to the disk fraction, especially at
older ages.

Other discussions of inner disk lifetimes have used different techniques and
more limited samples of stars (e.g. Walter et al. 1988, Strom et al. 1989, 
Skrutskie et al. 1990, Beckwith et al. 1990, Strom 1995; Haisch et al. 2001).
However, the general conclusions regarding inner disk lifetimes in the 3-10
Myr age range are, broadly speaking, similar to our findings of $<$2-3 Myr.
In addition, Muzerolle et al. (2000) 
demonstrate an indirect correlation of {\it accretion rates} with age, 
with at least several stars
showing measurable accretion signatures beyond 10 Myr.
So although most disks appear to evolve relatively rapidly, a small 
percentage may retain proto-planeary
nebular material for factors of 5-10 longer than does the average disk.

\subsection{Outer Disks}

Mid-infrared wavelengths, $\sim$10-90 $\mu$m,   
probe disk radii $\sim$1-5 AU, equivalent to 
the outer terrestrial and inner gas giant planetary zones of our solar system.
To date, observational sensitivity has been the primary hindrance to 
measurement of outer circumstellar disk evolution.  Data from the IRAS and ISO 
satellites, when presented in the same form as Figure 2 (e.g. Meyer \& Beckwith 2000;
Robberto et al. 1999), show similar morphology with $\sim$10 Myr needed for
depletion of 90\% of optically thick outer disks. 
The implication is that outer disk dissipation times are only slightly longer
than, or perhaps even consistent with, inner disk dissipation times, and hence 
that disk evolution is both rapid and relatively independent of radius. 
Spangler et al. (2001) and  Habing et al. (2001), however, argue for
a much longer time scale, on the order of hundreds of Myr.  There may be 
some confusion in these studies between primordial and debris disks 
as a single, continuous evolutionary path is not expected.  As the dust
transitions from optically thick to optically thin, spectroscopy becomes
an especially important tool for tracing the mass in terms of
grain size distribution 
and composition.  Imminence of SIRTF, which is sensitive to 
{\it nearby stellar photospheres between 3.5 and at least 24 $\mu$m}
will enable statistical studies of primordial and debris disk
evolution on several AU scales.

Millimeter wavelength emission probes the outer ($\sim$50-100 AU) cold disk
regions and is also optically thin.  Most millimeter observations have been 
directed towards stars younger than $\sim$10$^7$ year, but because of the
distance of these populations, generally place only upper limits on dust 
masses beyond the optically thick disk phase (e.g. Duvert et al. 2000). 
Dust mass surveys of older (10$^7$ - 10$^9$ year), closer, candidate
debris disk stars (e.g. Zuckerman \& Becklin 1993, 
Jewitt 1994, Carpenter et al. 2003) also reveal mostly upper limits due
to current sensitivity limits, though also several detections with
dust masses as low as 10$^{-8}$ M$_\odot$ (still a factor of 
$\sim$10$^2$-10$^3$ above our own zodiacal dust disk).

\subsection{Length of Transition from Optically Thick to Optically Thin}

Once the process of disk dissipation starts, how long does it take
for an individual object to transition from optically thick to optically thin?
As argued by Skutskie et al. (1990) and Wolk \& Walter (1996) based on the
perceived
lack of ``transition" objects, the process of evolution from optically thick
to thin takes only a few hundred thousand years or less.
Similarly, Nordh et al. (1996) show 7-15 $\mu$m
flux ratios in Chamaeleon that are scattered around {\it either} the colors 
expected from flat/flared disks, {\it or} around photospheric
colors, with essentially no objects located in between these clusterings.
When does a particular system go from being primordial
(dominated by growth of smaller bodies into larger ones) to debris 
(dominated by destruction of larger bodies into smaller ones which
are then removed from the system via Poynting-Robertson drag + stellar wind 
effects)?
Figure 3, for example, shows the spectral energy distribution of a moderately
young ($\sim$30 Myr old) star which may be in the process of depleting its
primordial dust disk, or which may have begun regenerating a dust disk.
More detailed study, particularly dust spectroscopy, is needed in order to
begin answering these kinds of questions.

There {\it is} evidence for the growth of grains in young disks to sizes larger 
than are expected based on our knowledge of interstellar dust. Typically these
arguments have been made from measurement of the frequency dependence of 
opacity in the expression $\tau_{\nu}(r) = \kappa_{\nu} \times \Sigma (r)$ where
$\kappa_{\nu} \propto \nu^\beta$ and the $\beta=2$ appropriate for
interstellar dust often yields in measurements of optically thin 
sub-/millimeter spectral energy distributions to 
$\beta = 0-1$ (see Miyake \& Nakagawa, 1993).  
Calvet et al. (2002) discuss this and other evidence for grain
growth in the specific case of TW Hydra.  Expected consequences on spectra and 
overall spectral energy distributions are presented in a parameter study of
disk geometry and grain properties by D'Alessio et al. (2001).

\section{Other Parameters Affecting Disk Evolution: Companions}

As mentioned earlier, the orbital radii and eccentricities of those exo-solar 
planets discovered to date bear little resemblance to our own solar system's
planets. If the architecture of our solar system is indeed rare 
or even unique, it may be that variables other than disk mass, size, and
composition may be important. Environment, e.g.
location in a clustered vs. in an isolated star formation region could 
potentially influence the disk, as already discussed.
Stellar properties imparted from the cloud core or set during the collapse
phase and early evolution of the disk, such as mass, metallicity,
and specific angular momentum may play some role as well.
For the remainder of this section we focus on multiplicity and the
possible influence of stellar/substellar companions on disk evolution.

Although our Sun is not a member of a multiple star system, a significant
fraction (30-80\%) of all stars do appear to be born in binaries, triples,
or higher order multiples (see review by Mathieu et al. 2000).  
For solar-type systems in the solar neighborhood, the distribution of 
orbital periods
is gaussian in log and peaks at 180 days (Duquennoy \& Mayor 1991) or 30 AU
-- within our current solar system.  An intriguing study of binarity 
by Patience et al. (2002) showed a much lower peak in the semi-major axis 
distribution for stars which are members of clusters (4 AU) than for
less densely spaced young stars (40 AU, similar to the DM91 field star
distribution) and also a lower overall companion frequency (just 10\%).  
They surmise that the solar neighborhood field star 
distribution could be comprised of star-forming regions that were 
$\sim$70\% clusters and $\sim$30\% low density regions, 
i.e. that {\it most} stars form with companions within 2-8 AU.

What is the role of multiples in disk evolution and are there implications 
for planet formation? Potentially planet-building circumstellar disks do exist 
around young multiples (e.g. the GG Tau ``prototype") as well as singles. 
And planets do form in widely separated binary systems (e.g. Cochran et al. 1997; Lowrance et al. 2002).  Little is known, however,
about the planet-forming potential of stars with high-mass brown dwarf and
stellar companions ($m_2/m_1$ $>$ 0.05) near the peaks of the semi-major axis
distributions, that is, with separations of only a few to several tens of AU.

Lubow \& Artymowicz (2000) calculate that
circumstellar disks around individual binary components will be 
truncated at radii in the range 0.2-0.5 times the orbital separation, while
any circumbinary disk surrounding the primary+secondary system will have an
inner radius of 2-3 times the orbital separation.  Although disks 
associated with binary systems are predicted to be truncated, they may still 
have sufficient surface density to form planets.  Lower mass (though not 
necessarily truncated) disks for 1-100 AU binaries 
are indeed suggested by observation (Jensen et al., 1996).

Several surveys capable of probing into the low-mass brown dwarf regime 
have already been conducted, enabled by
improvements in high-contrast imaging technology, but still limited in impact
by uncertainties involving physical association vs. chance projection.
Farihi et al. (2002) discuss an ongoing investigation of 
companions to white dwarfs, a survey which thusfar has revealed only
GD 165B (Becklin \& Zuckerman 1993).  
Schroeder et al. (2000) used HST/WFPC to survey stars within 13 pc, finding
no previously unidentified companions.  Lowrance (2001) 
investigates a sample of nearby GKM stars with HST/NICMOS, 
and finds several sub-stellar companions, 
while McCarthy (2001) uses coronographic techniques to study somewhat younger 
GKM stars, with a yield of no confirmed sub-stellar objects.  
Oppenheimer et al. (2001)
conducted a volume-limited survey of similarly mixed mass and age stars
finding only Gl 229B (Nakajima et al. 1995).  Generalizing the results of the
above investigations, sensitivities extend to masses of a few to tens of Jupiter 
masses over separations of a few tens to a few hundreds of AU.  
Wider companions 
($>$1000 AU) to FGKM stars are summarized by Gizis et al. (2001) and closer 
companions ($<$3-5 AU) are probed by doppler monitoring planet search programs.
Overall, the frequency of brown dwarf and planetary ($<$1-80 M$_{Jupiter}$) 
companions appears to increase
with orbital separation, rising from $<$0.3\% at $<$3 AU, to $<$3\% on tens
of AU scales, to $<$10\% by thousands of AU. 

Our own recently initiated survey can extend previous work in several 
directions.  Taking advantage of the fact that, at young ages,
the faintest of companions detectable with modern equipment are in fact 
giant planets, Metchev \& Hillenbrand (2002) describe adaptive optics 
coronographic work that is sensitive to companion masses as low as 
3-10 M$_{Jupiter}$ for the younger and closer portion of their sample.  
The parent stars are F8-K0 solar analogs 
(0.8-1.2 M$_\odot$), and the survey  explores a broad age range from 3 Myr
to 3 Gyr.  Complementing this work is an ongoing 
precision velocity program (Marcy et al.) that will reveal giant-planet and
any larger masses closer in to the central star than can be seen in direct 
or corongraphic imaging.  The target stars are
members of the Meyer et al. (2002) SIRTF/Legacy program for 
which dust disk information will soon be available, enabling a complementary
discussion of the effect of companions (stellar, brown dwarf, and planetary)
and disks. 

\section{Summary of Young Circumstellar Disks and Their Evolution}

Direct imaging has finally convinced even the most stubborn skeptics that 
circumstellar dust and gas {\it disks} do exist around young stars.
Detailed information on composition, size, mass, and even some hints regarding
surface density distributions are available in a limited number of cases.
However, we still need to rely on traditional photometric and spectroscopic 
techniques for the statistics required to understand disk dissipation 
time scales. At present, there is 
evidence for decreasing trends with age in: disk fraction, 
mean disk accretion rate, and mean disk mass. There are also signs in young
disks of evolution from interstellar grain parameters.  
What may be most interesting however, is the large dispersion about the mean 
for any given age, in all of these trends. 

By establishing over the next decade the decay with time of primordial dust via
near- and mid-infrared excess around
stars of different mass, we will have taken the first step in understanding
the possibilities for planetary formation.  Studies to determine the time 
scales for dust disk dissipation should be followed by those aiming to 
similarly quantify time scales for gas disk dissipation.  
Fully constraining the time period
over which raw materials needed for planetary formation are available means, 
ultimately, following the evolution of disk surface density 
as a function of radius from the central star.  The outstanding problem 
in planning for this kind of statistically robust future is that we do not have
adequate samples of stars in the 5-50 Myr age range, a
critical time in planet formation and early solar system evolution. 

In the near-future, SIRTF and SOFIA will provide an abundance of detail on
continuum spectral energy distributions and dust mineralogy, as well as hints
regarding gas (H$_2$) content.  Further afield, NGST and CELT will study gas 
disk evolution with the same rigor we are about to study dust disk evolution.
ALMA will provide images with the spatial resolution required for 
investigation of radial and azimuthal disk structure before and  
during the planet-building phase.  Early investigations (such as those 
currently proceeding)
of the influence of stellar and sub-stellar companions on disk characteristics
will be supplanted by those using improved ground- and space-based 
high contrast imaging 
and interferometric (e.g. Keck-I and SIM) technology. 
Exo-solar planet statistics continue to assemble. 

The connection between disks and planets is increasingly obvious.

\newpage

\begin{figure}[t]
\vskip-1.5truein
\plotone{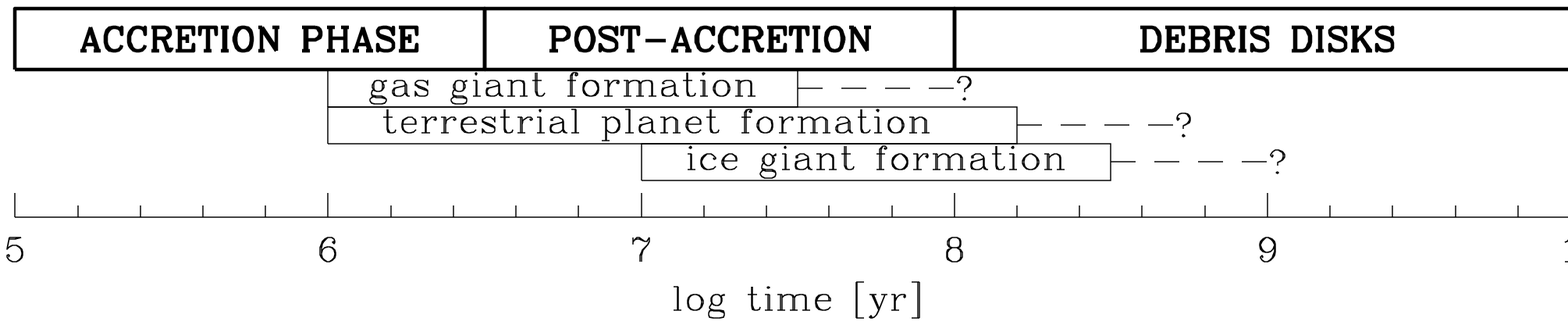}
\caption{Images of disks at various evolutionary stages
scaled to a timeline showing our general understanding 
of the basic phenomena.
Data are courtesy of
J. Stauffer and B. Patten (left panel, Ori 114-426 optically thick
``silhouette disk'' as seen with HST/WFPC), 
Kalas \& Jewitt 1995 (middle panel, $\beta$ Pic as seen in ground-based
coronographic imaging), and
P. Kalas (right panel, our own zodiacal dust disk along with a comet, as
photographed from Calar Alto).}
\end{figure}

\begin{figure}[h]
\plottwo{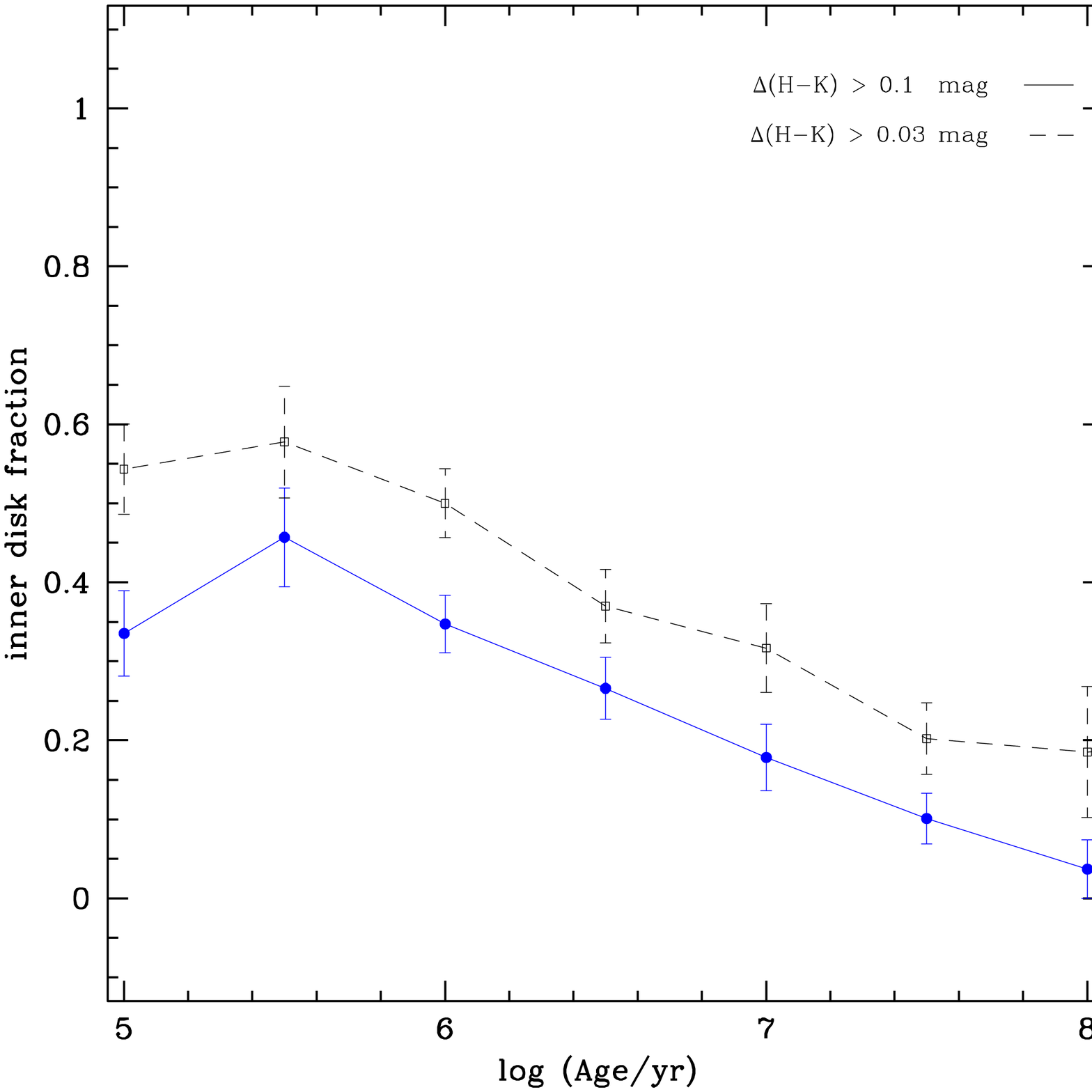}{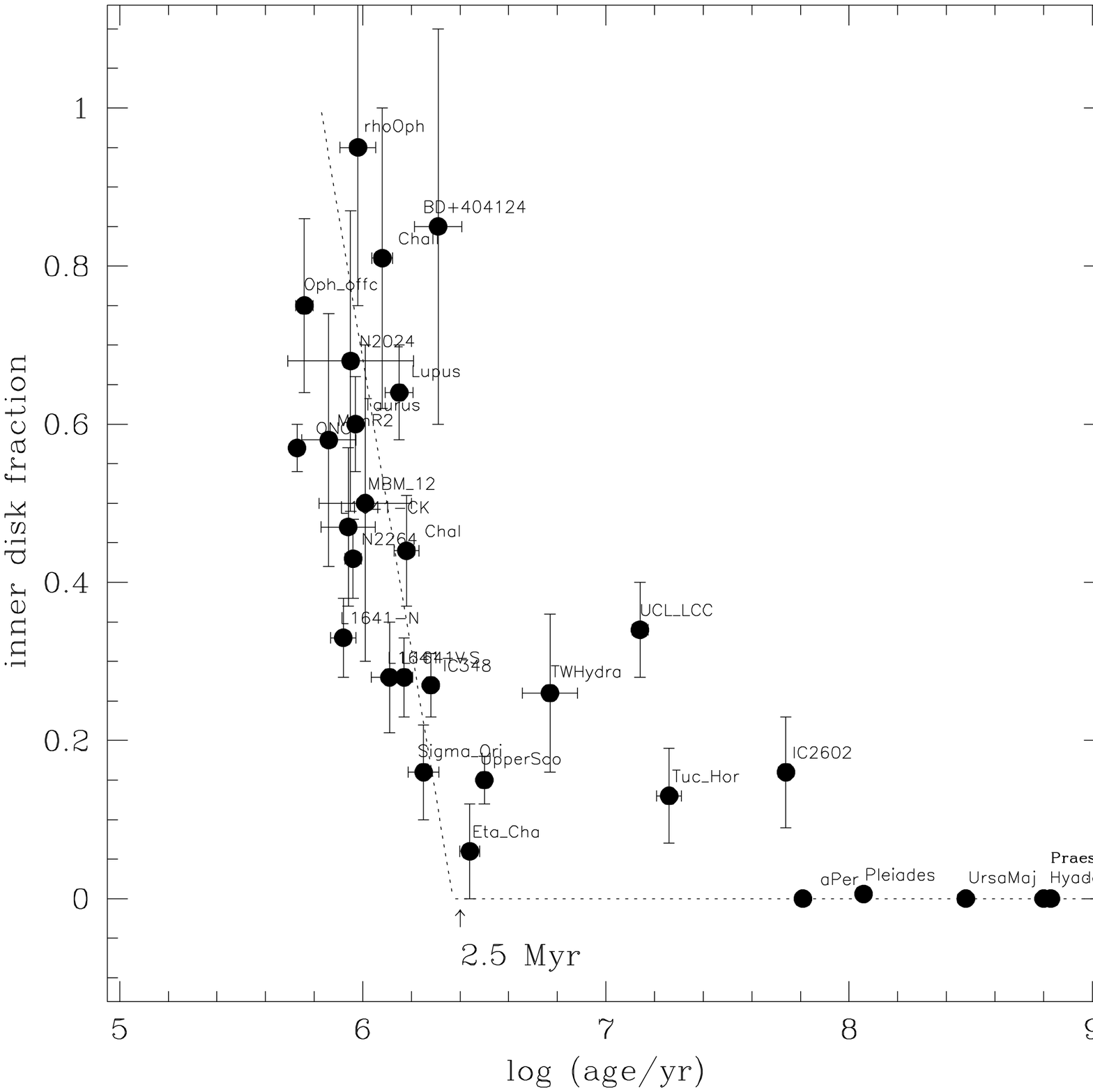}
\caption{
  {\it Inner accretion disk fraction vs. stellar age} inferred from
  H-K excess measurements for $\sim$3000 stars. In the left panel, each star
  contributes individually to the age bin found from its location on the HR
  diagram.
  The dashed line uses a cut of $\Delta$(H-K) $>$ 0.03 mag to define a disk
  while the solid line uses the more conservative $\Delta$(H-K) $>$ 0.1 mag.  
  Note that non-gaussian errors tend to scatter stars into older
  age bins, thus inflating the disk fraction at older ages.
  In the right panel, clusters are treated as single units of age corresponding
  to the median age found from the HR diagram.  Here, an intermediate cut of 
  $\Delta$(H-K) $>$ 0.05 mag is used to define a disk.  
  }
\end{figure}

\begin{figure}
\vskip-2.8truein
\plotone{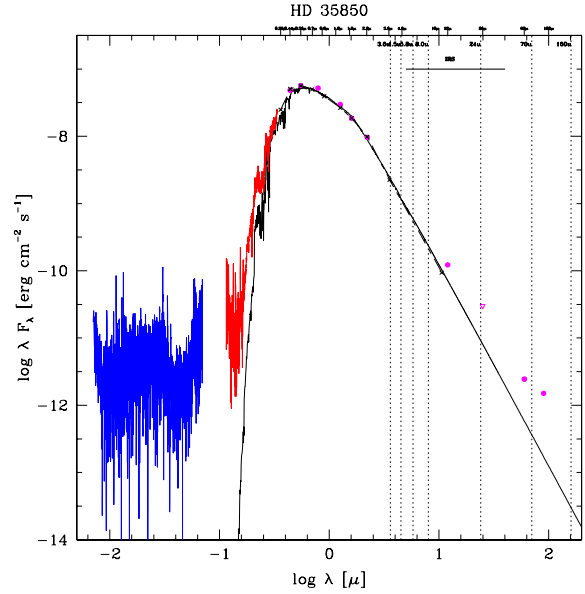}
\vskip 0.3truein
\caption{
Observed spectral energy distribution of HD 35850 (F8 V star at 27 pc)
from extreme-ultraviolet (EUVE; IUE) through optical/infrared 
(Hipparcos; 2MASS) to far-infrared (IRAS; ISO) wavelengths, compared to
a Kurucz model atmosphere. This star is thought to be between 10 and 100 Myr 
old and has either a late-stage primordial disk or an early-stage
debris disk. Vertical dotted lines indicate effective wavelengths
of SIRTF filters.
}
\end{figure}

\end{document}